\documentstyle[sprocl]{article}

\def\Journal#1#2#3#4{{#1} {\bf #2}, #3 (#4)}

\begin{document}
\title{STRUCTURAL INVARIANCE: A LINK BETWEEN CHAOS AND RANDOM MATRICES}
\author{F. LEYVRAZ, T.H. SELIGMAN}
\address{Laboratorio de Cuernavaca, Instituto de F{\'\i}sica,
University of Mexico, Cuernavaca, Morelos, MEXICO}
\maketitle\abstracts{
The concept of structural invariance previously introduced by the
authors is used to argue that the connection between random
matrix theory and quantum systems with a chaotic classical counterpart
is in fact largely exact in the semiclassical limit, holding for
all correlation functions and all energy ranges. This goes considerably
further than the usual results obtained through periodic orbit
theory.These results hold for eigenvalues of bounded time-independent
systems as well as
for eigenphases of periodically kicked systems and
scattering systems.}
\def\part#1#2{{\partial #1\over\partial#2}}
\def\ggg{{\cal G}}
\def\group{\ggg\times\ggg}
\def\hhh{{\cal H}}
\def\jpa#1#2#3{J. Phys. A: Math. Gen. {\bf{#1}}, #2 (19#3)}
\def\jpl#1#2#3{Jour. Physique Lettres {\bf{#1}}, #2 (19#3)}
\def\jmp#1#2#3{J. Math. Phys. {\bf{#1}}, #2 (19#3)}
\def\prl#1#2#3{Phys. Rev. Lett. {\bf{#1}}, #2 (19#3)}
\def\pra#1#2#3{Phys. Rev. A{\bf{#1}}, #2 (19#3)}
\def\prb#1#2#3{Phys. Rev. B{\bf{#1}}, #2 (19#3)}
\def\prc#1#2#3{Phys. Rev. C{\bf{#1}}, #2 (19#3)}
\def\prd#1#2#3{Phys. Rev. D{\bf{#1}}, #2 (19#3)}
\section{Introduction}%
The purpose of this paper is to establish a firm basis for a well-known
association beween random matrix theory on the one hand and quantum
systems with a chaotic classical counterpart on the other.
The key concept in our
approach will be that of structural
invariance already introduced in a previous paper
\cite{struct,selig}. Since this concept is
at the very heart of the issue, we start by describing it in the most
general terms possible. We then return to our specific purpose later.

The point at issue is the construction of
a reasonable ensemble given an individual system. Such problems are not
really new to physics: indeed, whenever one attempts to describe
an individual system by an ensemble, as occurs for example in statistical
mechanics, the problem of deriving the ensemble from the
individual system arises.
In such cases it is well known that one
first needs to identify a set of relevant properties. Once this
has been done, we define the structural invariance group of the
object as the group of all transformations which transform the
given object into one with the same relevant properties. To fix ideas, let us
consider first the following example: Take an arbitrary binary sequence
of $N$ elements. Assume first
that this sequence has no specific properties that
``strike the eye''.  The structural invariance group then consists of
all maps which carry binary sequences into other binary sequences, without
any restrictions on these maps.
On the other hand, assume that the sequence contains,
say, seventy percent ones. This cannot be chance, at least for $N\gg1$. We
therefore must limit the structural invariance group to those transformations
which respect that property. The structural invariance group is therefore
the group of all permutations acting on the binary sequence. If further,
say, the arrangement is periodic with period $r$, then the
structural invariance group will only consist of those permutations
which respect that feature and will therefore be isomorphic to the
symmetric group over $r$ elements. Note carefully how the structural
invariance group becomes {\it smaller\/} as the symmetry of
the object increases. Indeed, it may well be that
the structural invariance group is in a technical sense
complementary \cite{marcos,selig}
to the symmetry group with respect to the largest
group of all possible transformations.
As one readily sees in the above example,
the concept of structural invariance group is somewhat ambiguous in pure
theory, yet in practice the assignment of a given group to
an object hardly ever presents real problems.

Once the structural invariance group $G$ of an object has been defined, there
is usually no difficulty in embedding that object in a ``natural''
ensemble. The procedure is as follows: let the group $G$ act on the original
object. In this way it defines a set $\Sigma$, on which an action of $G$ is
defined. If $G$ has further an invariant measure, then this measure
induces a measure on $\Sigma$, so that we have indeed constructed
an ensemble of which the original object is a typical element.
If we now study such properties of the ensemble as are {\it ergodic\/},
that is, which hold with probability one for any given element of
the ensemble, we can reasonably conclude that the original object
must have these properties. If it does not, this strongly indicates that
the structural invariance group was incorrectly identified.

While these concepts may well be applicable in many different fields,
the authors were interested in applying them to the following example:
ever since the pioneering papers of Berry \cite{berry} the
connection between the spectral properties of a quantum system
having a chaotic classical counterpart with those of
random matrix theory (RMT) has been a subject of intense
research. The result of many independent numerical tests has been
the following: as long as no symmetries are present (in particular
no discrete symmetries), the spectra of the resulting quantum system
has the behaviour of the appropiate matrix ensemble \cite{oriol,thomas}.
This means that it belongs to the universality class of the so-called
Gaussian Orthogonal Ensemble (GOE) if it is invariant under time
reversal and to that of the Gaussian Unitary Ensemble (GUE) otherwise.
On the other hand, if symmetries do exist, they are taken
into account by introducing independent ensembles for each
symmetry sector.
The validity of such modelling has been supported by a large quantity
of numerical work as well as by semiclassical considerations based
on the Gutzwiller sum formula
\cite{gutz,berry85}. Some apparent exceptions to this
association between chaos and random matrix behaviour are known
(e.g. the modular billiard on the surface of constant negative
curvature), but these have been found to be due to peculiar hidden
discrete symmetries of number-theoretical origin \cite{georgeot}.

In this paper, we shall explain this association in terms of the concept of
structural invariance outlined above. Such an approach cannot, by
itself alone, rigorously prove anything for a specific system.
Nevertheless, it provides a powerful instrument for justifying
the association of a given type of ensemble to a given class of systems.
Further, as we shall see, it provides results that periodic orbit
theory is (up to now) quite unable to show: in particular, it can
be shown that the spectral properties predicted by RMT hold exactly,
not only for large energy differences, as in the periodic orbit
approach. It can also be seen that arbitrary correlation functions
are correctly given by RMT, whereas this is a highly non-trivial
problem in periodic orbit theory. Thus, it can be said that the statements
we make are far stronger than those which can be shown by other means,
but that we are limited to saying that they are true ``generically''
or ``with probability one''. Even this,
however, may not be as severe a drawback as might at first appear:
Indeed, assume one were able rigorously to show that any chaotic system
without discrete symmetries had the appropiate RMT behaviour.
Since proving the absence of {\it
any\/} kind of symmetry whatsoever is an extremely
difficult task,
the usefulness of even such a strong
theorem is very limited indeed.  Therefore, precisely from
the very standpoint of mathematical rigor, such a theorem would be
useless. Of course, combined with physical intuition as to the
``likelihood'' of hidden symmetries, it becomes useful again, but at
that level our arguments are, we believe, equally compelling.

The rest of this paper is organized as follows: in Sec.~2 we discuss the
application of the structural invariance concepts described above to the
case of a canonical map without any structure. From this we obtain a well-known
connection between the eigenphases of a chaotic periodically time-dependent
system and the so-called Circular Unitary Ensemble defined by Cartan. In
Sec.~3 we show how various symmetries of the problem reduce the structural
invariance group and we show that the ensembles one is led to consider
are exactly those which are usually associated with such systems. In Sec.~4
we show how the results for eigenphases of maps generalize to the case of
Hamiltonian systems and their eigenvalues. In Sec.~5 we present
some comments and conclusions.
\section{Structural invariance for canonical maps}%
In the following, we shall consider bijective
canonical maps $C$ from a compact phase space $\Gamma$ onto itself. This might
represent a number of things, such as the Poincar\'e map of a bounded
time-independent system, a scattering map or the time evolution over one period
of a periodically driven time-dependent system. For simplicity
we shall usually think of it as the last of these.

We shall take the point of view that the relevant properties
of a canonical map are those which give rise to recognizable
organized structures upon iterating the map. This is rather natural in view of
the examples given above. Thus we are led to consider
invariant tori and cantori as part of the properties which the structural
invariance group must leave invariant. Further, such properties as the
existence
of discrete symmetries must of course also be preserved.
On the other hand the
exact location and properties of the isolated unstable
periodic orbits of the system are of course not relevant. Of further relevance,
though rather less trivial, is time-reversal invariance. Indeed, we say that
a map $C$ is time-reversal invariant (TRI) if there exists a non canonical
map $T$ with $T^2$ being the identity, such that
$CT=TC^{-1}.$
The classical example of this, of course, is the usual case where $T$
only changes the sign of the momenta but other cases are known as well.

Let us now consider a map $C$ which is wholly structureless. Then, defining the
group of all bijective canonical transformations to be $\ggg$, one finds that
the structural invariance group of $C$ is $\group$ with the following
action:
\begin{equation}
(S,S^\prime):C\longrightarrow SCS^\prime.\end{equation}
The set $\Sigma$ is then given as the set of all canonical transformations,
which is itself the group $\ggg$, and the action of $\group$ on $\ggg$
is given by Eq.~(1). The example therefore appears as an entirely trivial
one.

In making $\Sigma$ to an ensemble, however, we encounter a fundamental problem:
The groups $\ggg$ and $\group$ are both infinite-dimensional and
do not have a known, useful invariant measure. There is therefore no
natural way to define a measure on $\Sigma$ dictated by
invariance considerations alone.

On the other hand, we are interested in the consequences of the chaotic
nature of the map for quantum mechanics. In this case, a chaotic transformation
induces (via any of a large number of quantization procedures) a unitary map
from a Hilbert space onto itself. Since we are considering a compact
phase space, this Hilbert space is essentially finite dimensional
with dimension $N$ given by
$\vert\Gamma\vert/(2\pi\hbar)^f,$
where $\vert\Gamma\vert$ is the volume of $\Gamma$ and $f$ is the number
of degrees of
freedom in the system. If we are considering a Poincar\'e map, for example,
then $N$ as well as $\vert\Gamma\vert$ are related to the energy scale. Quite
generally, it can be argued that the semiclassical limit, which is the
only case we shall be concerned with, corresponds to $N\gg1$. In this
case we can disregard the complex issues concerning the ambiguities involved
in quantization and assign to every canonical transformation a
unitary transformation
as shown by Dirac\cite{dirac}
\cite{monroe}. Again up to effects which are outside of
the semiclassical approximation, this unitary map can be viewed as a map
on an $N$-dimensional space. That is, in the final analysis, we have mapped the
group $\ggg$ onto $U(N)$ within the appropiate approximations.

{}From this, however, our conclusion follows immediately: indeed, there is a
Haar
measure on $U(N)$ which is invariant under the left and right action of $U(N)$
on itself. We therefore reach the following conclusion: a completely
structureless map is to be associated with the ensemble of all unitary matrices
endowed with the corresponding Haar measure. This corresponds to saying
that the matrices are generic elements of the so-called Circular Unitary
Ensemble (CUE). This conclusion had been reached by entirely different means
for the specific cases of the scattering matrix and of periodically driven
systems. These methods employed semiclassical techniques
based on periodic orbit theory, which could
only show that the two-point function of the eigenphases corresponds with
the CUE result for large eigenphase differences. Our result implies
much more: it claims the same result for all energy ranges and for all
properties of the eigenphases. It also extends to eigenfunctions without any
problems, since our claim holds for the unitary matrix representing the
canonical map $C$. On the other hand, as we said in the introduction, it
is not capable of showing that a specific system has these properties with
absolute certainty. Rather, it very strongly suggests that if the properties
of the CUE are found to be violated, some significant structure must be present
in the map, which has been overlooked.
\section{Implications of various structural properties}%
In the following we discuss maps which are not wholly structureless and
show that they do indeed give rise to the ensembles by which they are
successfully described in RMT. Let us start by the most important
case of time-reversal invariance. We need to find the subgroup $\hhh$
of $\group$ which leaves the TRI property
unaltered. An easy calculation shows that \cite{struct}
\begin{equation}\hhh=\{(S,S^\prime):S^\prime=TS^{-1}T\}.\end{equation}
{}From this follows the corresponding quantum-mechanical result. Let us call
$U_C$
the unitary matrix corresponding to the canonical transformation $C$. Then the
transformations that correspond to those that are induced by $\hhh$ are the
following:
\begin{equation}U_S:C\longrightarrow U_SU_CU_TU_S^{-1}U_T,\end{equation}
where $U_T$ represents the {\it anti-unitary\/} matrix associated to $T$.
If $T$ can indeed be represented by complex conjugation (as we shall assume
henceforth, the opposite case being the one that gives rise to the symplectic
ensembles), then the action of $U_S$ on the {\it unitary symmetric\/}
matrix $U_C$ is given by $U_SU_CU_S^t$, where $^t$ denotes the transpose.
The set $\Sigma$ is now the set of all symmetric unitary matrices and
the measure $d\mu$ is the one that remains invariant under the action
defined by Eq.~(3). Again by a standard theorem, this measure
is unique and given by the so-called Circular Orthogonal Ensemble
(COE).

Let us now assume that the map $C$ has a given symmetry, say $P$. This
means that
$PC=CP.$
Under these circumstances, the subgroup of $\group$ which leaves the
symmetry invariant is given by
\begin{equation}
\hhh=\{(S,S^\prime):SP=PS;S^\prime P=PS^\prime\}.\end{equation}
Again, the set $\Sigma$ consists of all maps $C$ having the symmetry $P$.
As a subset of $U(N)$, it turns out to be the direct sum of all
the various symmetry sectors of $P$. If $P$ has non-trivial representations,
then degeneracy follows trivially, whereas the independence of distinct
symmetry sectors is equally clear. None of these results can be easily
derived in such generality from periodic orbit
theory. Also, we see
the reason why time-reversal invariance plays a role distinct
from other Abelian symmetries such as parity. Again this
is far from obvious in a treatment based on periodic orbit theory, since in
both cases one has systematically doubly degenerate orbits.

Let us now consider a somewhat more involved case: Let us aasume that
we have a system with time-reversal invariance {\it and\/} a discrete
symmetry of order higher than two. If we reduce the classical phase
space according to symmetry sectors, two cases can present themselves:
First, all symmetry sectors are separately TRI. In this case (which
has been the most common so far) the map can be described as an
uncorrelated superposition of COE's. On the other hand, it is equally
possible that time-reversal carries one of the symmetry sectors
into the other and vice versa. In this case it is easy to see
that $T$ must be represented by an antiunitary matrix which
interchanges the two sectors. This leads to a structural invariance
group with the following action on the two blocks
\begin{equation}(U_1,U_2):(U_{C,1},U_{C,2})\longrightarrow (U_1U_{C,1}U_2^t,
U_2U_{C,2}U_1^t),\end{equation}
where the indices denote the symmetry sector
on which the corresponding matrices operate. From this it follows
that the two sectors have degenerate eigenphases (Kramer's degeneracy)
and that these eigenphases obey CUE statistics.
\section{From eigenphases to eigenvalues}%
So far we have only treated the case of a canonical
map, the iteration of which gives the time evolution. An important
question, however, is to transfer this analysis to flows generated
by a time-independent Hamiltonian. The immediate problem with
this is that we must find a way to account for the fact that the
overall density of states is {\it not\/} given by that of the
corresponding matrix ensembles and can be rather arbitrary,
whereas the fluctuation properties are indeed given by the RMT
predictions. This difficulty was absent in the earlier cases since
the density of states was correctly predicted to be uniform.

To do this, we consider the energy-dependent Poincar\'e map
$C_E(p_s,q_s)$, where the subscript $s$ indicates that these
refer to phase space variables in the Poincar\'e surface of section.
If one denotes by $T(E)$ the corresponding unitary map, it follows
from the above that the eigenphases of $T(E)$ are distributed according
to some appropiate random matrix ensemble. We now need to carry this
information over to the eigenvalue distribution.

To this end one proceeds as follows: Bogomolny \cite{bog}
has shown that a
semiclassical quantization condition is the following: if $E$
is an eigenvalue, then
\begin{equation}\det({\bf1}-T(E))=0.\end{equation}
Thus, if the eigenphases of $T(E)$ are denoted by $\exp(i
\phi_j(E))$, then every time a given $\phi_j(E)$ goes through
zero, $E$ is an eigenvalue. It turns out that the whole procedure
is only semiclassical, as the map $T(E)$ is only unitary in the
semiclassical limit, but this is not a problem, since this limit is
in any case the only one we are able to handle. Also, outside
of the true semiclassical limit, the relation between chaos and
RMT is much more subtle: in particular, one has problems such as
(transient) Anderson localization, in which chaotic behaviour
and randomly distributed eigenvalues coincide. One also then has to deal
with finite tunneling probabilities and other phenomena associated
with the structure of our canonical map in the complex plane,
which we have left out of consideration entirely, as our understanding
of these is still very incomplete.

It therefore appears that we have reduced the problem of determining
the spectrum of a Hamiltonian $H(p,q)$ to the study of the energy
dependence of the eigenphases of the quantized version of its
Poincar\'e map. To handle this problem, we must first know
how the Poincar\'e map changes under infinitesimal changes of
$E$. To this end, let us consider $S(E;q_s,q_s^\prime)$, defined
as the action along the classical path connecting the two surface
points at energy $E$. It follows from standard arguments that this is
in fact the generating function of $C_E(p_s,q_s)$. From
this one obtains after some manipulation
\begin{equation}C_{E+\Delta E}C_E^{-1}=\Phi_{\Delta E},
\end{equation}
where $\Phi_{\Delta E}$ is defined as follows: consider the ``Hamiltonian''
${\cal T}_E(p_s,q_s)$, which is defined as the time necessary to return
to the Poincar\'e surface if one starts from $(p_s,q_s)$. This ``Hamiltonian''
generates a flow on the Poincar\'e surface and $\Phi_{\Delta E}$ is the
infinitesimal canonical transformation corresponding to following this flow for
a ``time'' $\Delta E$.

If we now follow this through the quantization procedure, we obtain the
following: let the eigenphases $\phi_j(E)$ be defined as previously. Let
$\psi_j(E)$ be the corresponding eigenfunctions. If we now denote
by ${\cal H}_E$ the self-adjoint operator corresponding to ${\cal T}_E$, we
finally obtain
\begin{equation}{d\phi_j\over
dE}=\langle\psi_j(E)\vert\hhh_E\vert\psi_j(E)\rangle.\end{equation}
Now we must make some key approximations: First, we remmember that we are
in the semiclassical limit, that is, that the classical function
${\cal T}_E(p_s,q_s)$ is smooth compared with the $\psi_j(E)$. This
implies that instead of using, say, the Wigner distribution in
computing the r.h.s of Eq.~(8), we can use a smoothed version such as the
Husimi
distribution without great error. Since the $\psi_j(E)$ are eigenfunctions
of a matrix representing a totally structureless map, their Husimi
distributions
will be spread uniformly all over phase space. This would follow from our
considerations on structural invariance, but is equally confirmed
by a rigorous theorem of Shnirelman's concerning eigenfunctions
of chaotic billiards \cite{shnirelman}. From this one finally gets
\begin{equation}
{d\phi_j\over dE}=\overline{{\cal T}_E(p,q)},\end{equation}
where the overline denotes average over phase space. The crucial points to note
are the following: First, the velocity at which the eigenphases move is,
in a first approximation, independent of $j$. This is because there are $N$
eigenphases on the unit circle. Since they move with a velocity on the order
of one, they will cross zero at energies which differ by an order of $1/N$,
which is natural, since we have chosen our scale of energies to be the
classical one. Therefore, from one eigenvalue to the next, the velocity at
which the eigenphases moves hardly changes. This means that the CUE (or
other RMT) properties of the eigenphases translate directly into
corresponding properties of the eigenvalues of the system. Second, however, and
equally important: this constancy does not hold forever. Two effects
will eventually alter this state of affairs: first, the average on the
l.h.s. of Eq.~(9) will experience a secular change as $E$ changes. This
corresponds to the secular change in the density of states which is usually
eliminated by unfolding the spectrum. On the other hand, another effect may
well come into play even before this secular change becomes noticeable. The
point is that Eq.~(9){} is only true as a statistical statement and there are
fluctuations around the mean velocity. The most obvious cause for fluctuations
are departures of the Husimi distribution from equidistribution. Such
deviations
are well-known to exist, namely the so-called ``scars'' near short periodic
orbits. It could therefore well be that these accumulated fluctuations could
account for some of the effects due to short periodic orbits. To explore this
possibility, however, we would presumably require an understanding of
scars which we do not have at present. Finally, it should be pointed out that
the l.h.s of Eq.(9){} is directly related to the phase space volume
(see e.g. \cite{bog}) and
hence to the Weyl formula, so that this formalism recovers well-known results
on the overall density of states.
\section{Conclusions}%
We have presented a systematic approach to the connection between
RMT and individual dynamical systems. This connection is in a sense
of a probabilist type: it rests basically on conclusions of the form:
a given system is a typical representative of a certain ensemble,
elements of this ensemble have a given property with probability one,
therefore the original system has the stated property. While this
line of argument is wide open to fundamentalist attacks from the
mathematical side, it is undeniably useful
at the heuristic level. Further, similar lines of reasoning
are frequently used in statistical mechanics
when applying an ensemble description to an individual system. A
more genuine concern concerns the construction of the
ensembles: as we are not able to construct ensembles on the set of
all canonical transformations, we must first construct a set of
canonical
transformations at the purely classical level and then translate
this into quantum mechanics in order to obtain a reasonable candidate
for a measure. This is undoubtedly unsatisfactory, but it is
probably a genuine problem, not merely a measure of the authors'
incompetence. In fact, finding a connection between the approach
based on periodic orbits
and the one we have sketched here is a very interesting
open problem. In particular, periodic orbit theory, in order to be
consistent witht RMT, strongly suggests that peculiar correlations
between periodic orbits must exist for very long orbits
\cite{uzi}. Deriving
such results from considerations of structural invariance might
indeed be a significant step forward. That this is not
trivial, however, depends precisely on the fact that it is not
possible to obtain measures on the set of all canonical maps.
Similar remarks obtain for quantum localization:
as long as the phase space $\Gamma$ is compact,
localization is only transient, and therefore outside the
immediate range of application of semiclassics. Non-compact
phase spaces, on the other hand, also present problems
relating to the existence of an invariant measure.

On the other hand, the method of structural invariance has
shown itself to be a powerful tool: it associates the correct
ensemble in every relevant situation.
Further,
our approach shows that not only do the RMT predictions hold for the two-point
function and at large energy distances, but that they should
hold at all energy scales and for all correlation functions. This
is much more than periodic orbit analysis has achieved yet. In fact
the ``unreasonable effectiveness'' of RMT
in describing the quantum analogs of classically chaotic systems has so
far eluded any explanation. The result presented here
should therefore be viewed as a new and significant one, although entirely
expected from  the numerical point of view.

\end{document}